\documentclass{physeauth}

\usepackage{graphicx}
\usepackage{amsmath}
\usepackage{amssymb}

\begin{document}

\begin{frontmatter}

\title{Multiple-quasiparticle agglomerates at $\nu=2/5$} 

\author[difi]{D. Ferraro},
\author[lamia]{A. Braggio \thanksref{thank1}},
\author[difi]{N. Magnoli},
\author[lamia]{M. Sassetti}

\address[difi]{Dipartimento di Fisica \& INFN, Universit\`a di Genova,Via Dodecaneso 33, 16146, Genova, Italy}
\address[lamia]{LAMIA-INFM-CNR \& Dipartimento di Fisica, Universit\`a di Genova, Via Dodecaneso 33, 16146, Genova, Italy}

\thanks[thank1]{
Corresponding author. 
E-mail: braggio@fisica.unige.it}

\begin{abstract}

  We investigate the dynamics of quasiparticle agglomerates in edge
  states of the Jain sequence for $\nu=2/5$. Comparison of the  Fradkin-Lopez
model with the Wen one is presented within a field
  theoretical construction, focusing on similarities and 
  differences.  We demonstrate that both models predict the same
  universal role for the multiple-quasiparticle agglomerates that
  dominate on single quasiparticles at low energy.  This result is
  induced by the presence of neutral modes with finite velocity and is
  essential to explain the anomalous behavior of tunneling conductance
  and noise through a point contact.  

\end{abstract}

\end{frontmatter}

%%%%%%%%%%%%

\section{Introduction}

Noise experiments in point contacts have been crucial to demonstrate
the existence of fractionally charged quasiparticles in fractional
quantum Hall systems~\cite{Laughlin83}. In particular, it was proved
that for filling factor $\nu=p/(2np+1)$, with $n,p\in\mathbb{N}$,
(Jain series [2]), the quasiparticle (qp) charge is given by
$e^*=e/(2np+1)$ \cite{dePicciotto97,dePicciotto97-2,dePicciotto97-3}.  
A suitable framework for the
description of these phenomena is provided by the theory of edge
states~\cite{Wen90,Wen91}. For the Laughlin series ($p=1$) a chiral
Luttinger Liquid theory ($\chi$LL) with a single mode was proposed and
shot-noise signatures of fractional charge were
observed~\cite{dePicciotto97,Kane94}. For the Jain series ($p > 1$)
extensions were introduced either by considering $p-1$ additional
hierarchical fields, propagating with finite velocity (Wen 
model~\cite{Wen92}), or by considering two fields, one charged and one
neutral and topological (Fradkin-Lopez model~\cite{Lopez99,Chamon07}).
Recently, transport experiments performed on a point contact at
extremely low temperatures have shown unexpected change in the
temperature power-law of the conductance and tunneling particles with
a charge that can reach $p$ times the single quasiparticle
charge~\cite{Heiblum03,Heiblum03PE}.  
In~\cite{Ferraro08} we
proposed a possible explanation of these effects by introducing a
generalized Fradkin-Lopez model (GFL) with neutral modes propagating at
finite velocity.  This assumption was crucial in order to lead the
agglomerates of qps to dominate the tunneling processes\cite{Ferraro08}.

In this paper we will address the question about the robustness of
this important result. For this reason we will compare the above
mentioned GFL model with the one proposed by Wen~\cite{Wen92}.  We
will focus on filling factor $\nu=2/5$, which is one of the available
case considered in experiments.  We will prove that both models
predict the same universal role for the qp agglomerates, and that are
able to explain, with similar level of accuracy, the experimental
observations.  These facts further support our interpretation of
experiments, confirming the importance of agglomerates in tunneling
processes for the Jain series~\cite{Ferraro08}.

\section{Quasiparticle agglomerate field construction}
\label{sec:Multipleqp}

Let us start this section with some general remarks.\newline  The bulk
excitation wave functions in the Hall fluid have to satisfy the
no-monodromy requirement~\cite{Froehlich97,Ino98} and have to be of
single-valuedness with respect to the electrons.  This means that the
phase acquired by any excitation in a loop around an electron must be
a multiple of $2\pi$.  Considering edge excitations, the "holographic
principle" shows that any bulk excitation can be mapped into operators
defined at the boundary~\cite{Susskind95}. 
So the above no-monodromy condition, for an
excitation at the edge, can be expressed in terms of constraint on the
mutual statistical angle between the excitation itself and the
electron.  In general, the mutual statistical angle $\Theta$ between two edge operators
$\Psi(x)$ and $\Psi'(x)$ is defined as
\begin{equation}
\label{eq:mutualstatistics}
\Psi(x) \Psi'(x')= \Psi'(x') \Psi(x)e^{-i\  \Theta\  \textrm{sgn}(x-x')}.
\end{equation}
The no-monodromy condition requires that the mutual statistical angle
between any quasiparticles operator and the electronic operator must be an
integer multiple of $\pi$. Note that $\Theta$ corresponds to the usual definition of the
statistical angle $\theta$ if the two operators
in~(\ref{eq:mutualstatistics}) coincide~\cite{Wilczek90}.

In the following, the field representation of edge excitations will be
obtained along two main steps. First, one has to identify the possible
electron operators, which in the end will form the electron field. 
They must have unit charge $e$, fermionic statistics 
and they have to mutually satisfy the no-monodromy condition.\newline 
Second, one has to identify the operators of the
qps and of the agglomerates with the appropriate charge, 
statistics and no-monodromy requirement.

We will classify all the excitations in terms of their charge 
that has to be an integer multiple of the quasiparticle (qp) charge $e^*=e/5$. We will 
refer to an $m$-agglomerate as an excitation with a charge $me^*$. Note that 
the electron is the agglomerate with $m=5$.

\subsection{General theoretical framework at $\nu=2/5$}

We start with the description of a single infinite edge at $\nu=2/5$.
Both  GFL and Wen models are described in terms of  two bosonic
fields.{\footnote{The original Fradkin-Lopez theory postulates two
  neutral modes but, for infinite edges, it is possible to
  consider an ``effective'' theory with a single neutral
  mode~\cite{Chamon07}.)}
Here, we do not consider the details of the derivation, but we refer to 
the available literature~\cite{Wen90,Wen91,Lopez99,Chamon07,Ferraro08}. 

The edge consists of a charged mode $\phi^+$ and a neutral mode
$\phi^-$, mutually commuting.  
 The real-time action $\mathcal{S}$ is ($\hbar=1)$
\begin{eqnarray}
\label{eq:action}
\mathcal{S}=&&\frac{1}{4\pi\nu_{+}}\int\!\! dt dx\  \partial_x \phi^+
(-\eta^{+}\partial_t-v_+\partial_x)\phi^++\nonumber \\
&&\frac{1}{4\pi \nu_{-}}\int\!\! dt dx\  \partial_x \phi^- 
(-\eta^{-}\partial_t-v_-\partial_x)\phi^-\ ,
\end{eqnarray}
with  $v_\pm$ the charge and neutral mode velocities.
The charge parameters are common in both models with $\eta^{+}=1$ and $\nu_{+}=2/5$, while the coefficients associated
to the neutral mode are model dependent~\cite{Lopez99,Ferraro08,Wen03}
\begin{eqnarray}
\label{GFL}
&&\eta^{-}=-1\qquad \nu_{-}=1\qquad {\rm GFL~~model}\\
&&\eta^{-}=1 \qquad ~~\nu_{-}=2\qquad {\rm Wen~~ model}\,.
\label{wen}
\end{eqnarray}
One can see that the neutral mode is counter-propagating in the GFL model and
co-propagating in the Wen one.  The electron number density is
$\rho(x)=\partial_x \phi^+(x)/2\pi$ and the commutation
rules among different fields are $[\phi^{\pm}(x),\phi^{\pm}(x')]=i \pi \eta^{\pm}
\nu_{\pm} \textrm{sgn}(x-x')$.\newline  Using the bosonization technique the 
edge operators are expressed as exponential combinations of the bosonic
edge fields.  We start with the operator associated to the
$m$-agglomerate~\cite{Wen03}
\begin{equation}
\label{eq:noperator}
\Psi^{(m)}(x)=\frac{1}{\sqrt{2\pi a}}e^{i [\alpha_m
\phi^+(x)+\beta_m\phi^-(x)]}\ ,
\end{equation}
here, $a$ is the short cut-off
length and $\alpha_m$, $\beta_m$ represent the coefficients of the
charged and neutral modes respectively. The pair of their  values  
$(\alpha_m,\beta_m)$  determines uniquely an edge excitation
and it will be used as an alternative notation for the operator
$\Psi^{(m)}(x)$.

The physical properties of the operator~(\ref{eq:noperator}) can be
now obtained as follows. The charge coefficient is derived using
the commutation with the electron density
\begin{equation}
\label{eq:charge}
[\rho(x),\Psi^{(m)}(x')]=-Q_m\delta(x-x')\Psi^{(m)}(x')\ , 
\end{equation}
with $Q_m=me^*/e$ the $m$-agglomerate charge in unit of $e$. Applying
the commutations among the bosonic fields one has 
\begin{equation}
\label{eq:chargeQm}
\alpha_m=\frac{Q_m}{\nu_+}=\frac{m}{2}\ .
\end{equation}
Statistical properties between different operators are characterized
by the mutual statistical angle in
Eq.(\ref{eq:mutualstatistics}). This angle depends on the coefficients
$(\alpha,\beta)$ and $(\alpha',\beta')$ that define two different
operators.  By using the commutation rules  one obtains
\begin{equation}
\label{eq:thetaMut}
\Theta\left[(\alpha,\beta),(\alpha',\beta')\right]
=\pi\left(\eta_+\nu_+\alpha\alpha'+\eta_-\nu_-\beta\beta'\right)\,.
\end{equation}
Statistical properties among equal operators are
defined in terms of the statistical angle $\theta_m$ which 
is fixed in the Chern-Simons
effective theory of hierarchical fractional quantum Hall states. For
the $\nu=2/5$ it is~\cite{Lopez99,Ferraro08}
\begin{equation}
\label{eq:theta}
\theta_m=-\pi m^2\left(\frac{7}{5}\right) - 2\pi k,
\end{equation}
with $k\in\mathbb{Z}$, encoding the $2\pi$ periodicity. 
 For  
the $m$-agglomerate in~(\ref{eq:noperator}) the explicit  
expression in terms of the pair $(\alpha_m,\beta_m)$ is
\begin{equation}
\label{eq:thetaAB}
\theta_m=\pi\left[\nu_+(\alpha_m)^2+\eta_-\nu_-(\beta_m)^2\right]\,.
\end{equation}
Note that once the charge of the agglomerate is fixed, 
one can still choose operators differing in the statistical angle
by an integer of $2\pi$. 
If they satisfy the no-monodromy condition with electrons they are indeed
admissible operators.

Hereafter we will apply these general results to the Fradkin-Lopez model in
order to derive the set of admissible operators for the agglomerates.

\subsection{Fradkin-Lopez model}
Let us start to identify the possible electron operators that
correspond to an agglomerate with $m=5$. The parameters for the
Fradkin-Lopez model are in Eq.(\ref{GFL}). The electron charge fixes
the charge coefficient~(\ref{eq:chargeQm}) to be
$\alpha_e=5/2$~\footnote{ For simplicity of notation we will use from
  now the subindex $e$ to denote the electrons, instead of $m=5$.}.
The possible values of $\beta_e$ will be determined, first of all, by
comparing the expressions (\ref{eq:theta}) and (\ref{eq:thetaAB}).
This constrain restricts the possible solutions\footnote{Note that we
  can freely redefine $k$ making use of the $2\pi$ periodicity.} to
$\beta_e(k)=\pm\sqrt{3/2+2k}$ with $k\in\mathbb{N}$. In addition, one
has still to impose the no-monodromy requirement with any other
electron operators
\begin{equation}
\label{eq:thetaMut1}
\Theta\left[(\alpha_e,\beta_e(k)),(\alpha_e,\beta_e(k'))\right]=\pi l
\end{equation}
with $l\in \mathbb{Z}$. Here, $k$ and $k'$ identify two different
electron fields.  By using the expressions~(\ref{eq:thetaMut}) and
(\ref{eq:thetaMut1}) and substituting the above values for
$\alpha_e,\beta_e(k),\beta_e(k')$ one obtains the condition for $k$ 
and $k'$
\begin{equation}
\label{kk'}
(3+4k)(3+4k')=(2h+1)^2
\end{equation}
with $h\in \mathbb{N}$. 
The more general solution in~(\ref{kk'}) can be represented in terms
of two integers $r$ and $q$ as $k_r[q]=r+(3+4r)(q^2+q)$.  Any
$r$-family defines an admissible set of electrons operators. In the FL
model one selects the $r=0$ family~\cite{Lopez99} with $k_0(q)=3q(q+1)$.  
In this case,  the neutral field coefficient is $\beta_{e}(q)=\sqrt{6}(q+1/2)$
with\footnote{Note that if $\beta(q)$ corresponds to an admissible
  value also $-\beta(q)$ will be admissible. We encode this property
  by admitting $q\in\mathbb{Z}$.}  $q\in\mathbb{Z}$.

It is important to note that the operators associated to $\beta_e(q)$
can be further classified into two sub-classes that differ only in the parity
of $q$: $\beta_{e,+}(q)$ for odd $q$ and $\beta_{e,-}(q)$ for $q$
even.  Within the same class the operators \emph{anticommute}, while
they \emph{commute} if they belong to two different classes, as can be
easily verified using Eq.(\ref{eq:thetaMut}). This last property defines a
parafermionic statistics \footnote{The condition of no-monodromy
  requires only that the mutual statistical angle must be integer and
  not an odd integer as the fermionic statistic would imply.}. 
In order to write the more general expression for the electron
operator one has to convert the parafermionic set of operators in a
fermionic one via a Klein transformation~\cite{Nambu74}. This is
achieved by defining two bosonic Klein factors for the even and odd
sub-classes respectively $\mathcal{F}^{(+)}$ and
$\mathcal{F}^{(-)}$ mutually anticommuting
$\{\mathcal{F}^{(+)},\mathcal{F}^{(-)}\}=0$.  It is easy to verify
that this restore the anticommutation properties between operators of
the two different classes. The more general electron operator can be then written as
a linear combination of all the possible electron representatives
\begin{equation}
\label{eq:electrFL}
\Psi_e(x)=\sum_{q\in\mathbb{Z}} c_q  
\mathcal{F}^{((-)^q)} e^{i [\frac{5}{2}\phi^+(x)+
\sqrt{6}(q+\frac{1}{2})\phi^-(x)]}\,
\end{equation}
where $c_q$ are the coefficients of the linear superposition.

Let us consider now the $m$-agglomerate with charge $me^*$ and 
$\alpha_m=m/2$ (cf. Eq.(\ref{eq:chargeQm})).  The neutral mode coefficient
$\beta_m(k)$ is again determined by the statistical angle $\theta_m$
given in Eq.(\ref{eq:theta}). One has
\begin{equation}
\beta_m(k)=\pm\sqrt{\frac{3}{2}m^2+2 k}\,,
\end{equation}
with $k\geq k^{\rm min}$ and $k^{\rm
  min}=-\mathrm{Int}[3m^2/4]$~\footnote{$\mathrm{Int}[x]$ denotes the integer
  part of $x$.}. As already pointed out the agglomerate operators
have to be also compatible with the no-monodromy condition with 
the electron operators of Eq.(\ref{eq:electrFL}). Imposing this
we find that, for a given $m$-agglomerate, $k$ is not an
arbitrary integer but it  assumes specific values only. In
particular, for odd agglomerates with $m=2s+1$, the admissible $k$ are
given by $k=3q^{2}+3q-3s^{2}-3s$ being $q\in \mathbb{N}$, while for
even agglomerates, $m=2s$, the possible values are
$k=3q^{2}-3s^{2}$. This implies the following values for $\beta_m(q)$
\begin{equation}
\beta_{2s+1}(q)=\sqrt{6}\left(q+\frac{1}{2}\right),\qquad
\beta_{2s}(q)=\sqrt{6} q
\label{beta}
\end{equation} 
with $q\in \mathbb{Z}$.  Note that the single qp and the electron
excitations belong to the odd  family with $(s=0)$ and $(s=2)$.  
Substituting the results for $\alpha_m$ and 
$\beta_m(q)$ in the operator expression~(\ref{eq:noperator})
we obtain the general form
\begin{equation}
\label{eq:psi_odd}
\Psi^{(2s+1)}(x)=\frac{1}{\sqrt{2\pi a}}e^{i(s+\frac{1}{2})\phi^{+}(x)+i 
\sqrt{6}\ (q+\frac{1}{2})\phi^{-}(x)}\ ,
\end{equation}
for odd agglomerates, and
\begin{equation}
\Psi^{(2s)}(x)=\frac{1}{\sqrt{2 \pi a}}e^{i s\phi^{+}(x)+i 
\sqrt{6}\  q\phi^{-}(x)}
\label{eq:psi_even}
\end{equation}
for even ones.

It is important to observe that the $5$-agglomerate family corresponds
exactly to the series of electron operators we found before in
Eq.(\ref{eq:electrFL}).  As a final remark we would like to comment on
the parafermionic statistics.  All the $m$-agglomerate with $m$ odd,
similarly to the electron case, can be divided in two classes.
Agglomerates within one class have the appropriate statistical angle
$\theta_m$ while agglomerates in different classes have the mutual
statistical angle equal to $\theta_m\pm\pi$.\newline The proper 
fractional statistics of the $m$-agglomerate can be then recovered by
using the bosonic Klein operators $\mathcal{F}^{(+)}$ and
$\mathcal{F}^{(-)}$ previously defined.
The more general operator of $m$-agglomerate will be then written as a 
linear combination of exponential operators similar to Eq.(\ref{eq:electrFL}).

To conclude we note the symmetries of the coefficients
$(\alpha,\beta)$ in Eqs.(\ref{eq:psi_odd}) and
(\ref{eq:psi_even}). When the charge coefficient $\alpha$ is
half-integer (integer) the neutral coefficient $\beta$ is always given
by $\sqrt{6}$ multiplied with an half-integer (integer).  These
numbers play a role analogous to the integers $m_{1}$ and $m_{2}$ that
appear in the original Fradkin-Lopez model~\cite{Lopez99}.

\subsection{Wen model}
Here, we would like to compare the results obtained for the GFL
model with the one deriving from the Wen model.  Field theoretical
description of the Wen theory was extensively treated in the
literature~\cite{Wen90,Wen91,Wen92,Kane95} and also presented in
textbooks ~\cite{Wen03}. We assume that the reader is familiar to the
basic results of the theory.

Wen showed that is possible to write all the physical operators as
$\Psi_{\mathbf{l}}(x)\propto e^{i[l_1\varphi_1(x)+l_2\varphi_2(x)]}$ 
where $\mathbf{l}=(l_1,l_2)$ is a vector with $l_1,l_2\in\mathbb{Z}$
and  $\varphi_r(x)$ are the edge bosonic fields with $r=1,2$ defined on 
the symmetric base. 
It is easy to show that these states correspond to the $m$-agglomerate 
fields obtained in the GFL model if we use the base of charged and neutral modes 
$\phi^{\pm}=\varphi_1\pm\varphi_2$. In this case the physical states 
are  $\Psi_{\mathbf{l}}(x)=e^{i[\frac{m}{2}\phi^+(x)+\frac{j}{2}\phi^-(x)]}$, 
where we introduced the two variables: $m=l_1+l_2$ as the number of qp in the 
excitation, and the $j=l_1-l_2$ that plays the role of neutral isospin. These two 
variables are integers and  have the same parity. For odd agglomerates with
 $m=2s+1$ and $j=2q+1$ the corresponding
operators ($q\in \mathbb{Z}$) are 
\begin{equation}
\Psi^{(2s+1)}(x)=\frac{1}{\sqrt{2\pi a}}e^{i(s+\frac{1}{2})\phi^{+}(x)+ 
i (q+\frac{1}{2})\phi^{-}(x)}\,.
\label{eq:psiW_odd}
\end{equation}
For even agglomerates  ($m=2s$ and $j=2q$) it is 
\begin{equation}
\Psi^{(2s)}(x)=\frac{1}{\sqrt{2 \pi a}}e^{i s \phi^{+}(x) + i q\phi^{-}(x)}\ ,
\label{eq:psiW_even}
\end{equation}
where we used the notation of Eq.(\ref{eq:psi_odd}) and 
Eq.(\ref{eq:psi_even}).

Note that the above field operators could have been obtained by following similar 
steps applied in the previous section for GFL model.\newline From
the comparison of Eq.(\ref{eq:psi_odd}) and Eq.(\ref{eq:psi_even})
with previous formulas we note that the GFL has exactly the same
operatorial structure of the Wen model. This indicates that the
electrons and all the $m$-agglomerates with odd $m$ have the peculiar
parastatistical properties found in the GFL model. Introducing again
the two Klein operators $\mathcal{F}^{(\pm)}$ we
can restore the correct statistical properties for all the
representatives and write the most general $m$-agglomerate as a linear
combination of all the operators with a final result very similar to Eq.(\ref{eq:electrFL}).

In conclusion, we observe that the difference between the GFL and the
Wen model at the level of $m$-agglomerate operators is present in the factor
$\sqrt{6}$ in the neutral coefficient. As we will see in the next section, this 
will play an important role in the evaluation of the scaling
properties.

\section{Scaling dimension of the agglomerates}

In this section we select, among the different excitations, the ones
that are dominant in tunneling processes.
We introduce the local scaling dimension $\Delta_m$ of the $m$-agglomerate, defined as 
half of the power-law exponent at long times
($|\tau|\to\infty$) in  the two-point
imaginary time Green function~\cite{Kane92} $\mathcal{G}_m(\tau)=\langle
T_\tau[\Psi^{(m)}(0,\tau)\Psi^{(m)\
  \dagger}(0,0)]\rangle\propto\tau^{-2 \Delta_m}$.  
For the operator in Eq.(\ref{eq:noperator}) it is at $T=0$
\begin{equation}
\label{eq:Gret}
\mathcal{G}_m(\tau)\propto\left(\frac{1}{1+\omega_+ |\tau|}\right)^{g_+
\nu_+ \alpha^2_m}\left(\frac{1}{1+\omega_- |\tau|}\right)^{g_-\nu_- \beta^2_m}\,.
\end{equation}
Here, $\omega_{\pm}=v_{\pm}/a$ represent the mode bandwidths setting the high energy cutoff.
The scaling dimension is then
\begin{equation}
\label{eq:scaling}
\Delta_m=\left[g_+\nu _+ (\alpha_m)^2 +g_- \nu _- (\beta_m)^2\right]/2\ .
\end{equation}
Note that the presence of the neutral mode contribution $\beta_m$ in
(\ref{eq:scaling}) is induced by a  finite bandwidth $\omega_-$ and
a finite velocity $v_-$.  This is the main generalization and difference with respect to 
the Fradkin-Lopez model which assumes topological neutral mode with $\omega_-=v_-=0$.

In order to take into account possible additional interaction effects
we considered in Eq. (\ref{eq:Gret}) renormalized parameters with
$g_{\pm}\geq 1$.  They correspond to the renormalization of the
dynamical exponents induced by a coupling of the fields with
independent dissipative baths~\cite{Rosenow02}.  The microscopic
models underlying these renormalizations were extensively treated in
literature~\cite{Rosenow02,Papa04-1,Papa04-2,Papa04-3} and will not be specifically
discussed here. Note that  the renormalizations do not affect the statistical
properties of the fields, which depend only on the equal-time
commutation relations, i.e. the field algebra.

Calculating the scaling dimension in Eq.(\ref{eq:scaling}) for the
operators in GFL model, described by
Eqs.(\ref{eq:psi_odd})-(\ref{eq:psi_even}) and in the Wen model with
Eqs.(\ref{eq:psiW_odd})-(\ref{eq:psiW_even}), it is easy to map the
two models for example choosing the substitution
\begin{equation}
\label{eq:mapping}
g_+^{W}= g_+^{GFL}\qquad g_-^{W}=3 g_-^{GFL}\ ,
\end{equation}
where $g_\pm^{GFL}$ and $g_\pm^W$ are respectively the generalized
Fradkin-Lopez and Wen $g_\pm$ parameters.  Having in mind this
relation we will now analyze the scaling behavior of the two models.

The \textit{most relevant} operator with the minimal scaling dimension 
will dominate the tunneling processes between two edges 
in the weak-backscattering limit~\cite{Ferraro08}. 
For a given $m$-family, the minimal value 
$\Delta^{\rm min}_{m}$ corresponds to the minimal value of  $(\beta_m(q))^2$ in Eq.(\ref{beta}),
this is for $q=0$.

In Fradkin-Lopez one has $\beta_{2s+1}(0)=\sqrt{3/2}$ and
$\beta_{2s}(0)=0$.  Agglomerates with $m> 2$ are never dominant
because the charge coefficient grows with $m$ and consequently from
Eq.(\ref{eq:scaling}) we have $\Delta^{\rm min}_{m>2}>\Delta^{\rm
  min}_{m=2}$. So we need to compare only the minimal scaling
dimension of the single qp $\Delta^{\rm min}_{1}$ with the
two-agglomerate $\Delta^{\rm min}_{2}$.  Simple calculations show that
the two qp agglomerates are \emph{always dominant} in the parameter
region $g_+^{GFL}/g_-^{GFL}<5$, otherwise the single qp tunneling
prevails~\cite{Ferraro08}. Using the mapping in Eq.(\ref{eq:mapping})
this means for the Wen model $g_+^W/g^{W}_-<5/3$. Note that the above
conditions are fulfilled in the case of unrenormalized parameters
($g_{\pm}^{GFL}=g_{\pm}^{W}=1$) for both models\footnote{In the
  unrenormalized Wen model the two qp agglomerate dominance at low
  energy was reported long time ago.~\cite{Kane95}}.  As a result the
scenario, previously used to explain the experimental
anomalies~\cite{Heiblum03,Ferraro08}, can be qualitatively applied to
both models.  From the above analysis we conclude that the
agglomerates are important excitations both for the Fradkin-Lopez and
the Wen models at $\nu=2/5$ in the weak back-scattering limit.

To summarize, we demonstrated that the family of operators are
essentially the same in the two models. The mapping of the scaling
dimension, demonstrating in both models the relevance of agglomerates
is in agreement with the recent scenario introduced to explain
experimental observations~\cite{Ferraro08}. These result suggest that
there is an urgent need of a renewed interest from the experimental
community in the quantum Hall system in the Jain series to better
clarify the agglomerate physics.

\section*{Acknowledgments} We thank M. Heiblum, E. Fradkin and S. Das
for useful discussions. A.B. acknowledges financial support by
 INFM-CNR via Seed Project PLASE001.

\end{document}